# Open Government Data Programs and Information Privacy Concerns: A Literature Review

Mehdi Barati

*ORCID Nr: 0000-0002-3447-4197*

*Department of Information Science, CEHC, State University of New York at Albany, New York, U.S.*
*mbarati@albany.edu*

*Abstract: This study presents a narrative review of the literature on privacy concerns of Open Government Data (OGD) programs and identifies suggested technical, procedural, and legal remedies. Peer-reviewed articles were identified and analysed from major bibliographic databases, including Web of Science, Digital ACM Library, IEEE Explore Digital Library, and Science Direct. Included articles focus on identifying individual information privacy concerns from the viewpoint of OGD stakeholders or providing solutions for mitigating concerns and risks. Papers that discussed and focused on general privacy issues or privacy concerns of open data in general or open science privacy concerns were excluded. Three streams of research were identified: 1) exploring privacy concerns and balance with OGD value propositions, 2) proposing solutions for mitigating privacy concerns, and 3) developing risk-based frameworks for the OGD program at different governmental levels. Findings suggest that contradictions with Fair Information Practices, reidentification risks, conflicts with OGD value propositions, and smart city data practices are significant privacy concerns in the literature. Proposed solutions include technical, legal, and procedural measures to mitigate privacy concerns. Building on the findings, practical implications and suggested future research directions are provided.*

*Acknowledgment: I am very grateful to Dr. Bahareh Ansari, Dr. Behlendorf, Dr. Tatar, and Dr. Yankson for their insightful comments.*

## 1. Introduction

Open Government Data (OGD) initiatives have been proposed and implemented by many governments as a tool to enhance transparency, accountability, efficiency, and effectiveness of government and public sector functions, public participation and collaboration among public, private, and civic organizations, and innovative creation of knowledge, services, and businesses (Ansari, 2021;

Francey & Mettler, 2021; Jetzek, 2016; Safarov et al., 2017; Wood et al., 2016). Government data is qualified as OGD if data is available, accessible, reusable, and redistributable (Ubaldi, 2013). This means that the released data must be available conveniently and free of charge to anyone, in a machine-processable format, and under terms and conditions that permit reuse and redistribution, including combining with other datasets. The ability to combine OGD with other private or public datasets by app developers to make better products and services is frequently cited as one of the economic goals of OGD initiatives in the literature (Ansari et al., 2022; Kjærgaard et al., 2020). In this review, all data that qualifies the openness criteria, produced by government and public agencies as a direct product or byproduct of service provision, and released to the public are considered OGD.

OGD barriers are studied intensively, including technical, organizational, legal, and usability barriers (Barati & Yankson, 2022; Crusoe & Melin, 2018; M. Janssen et al., 2012; Martin, 2014; Wang et al., 2019). Privacy and security concerns are discussed as legal and organizational barriers to the release and utilization of OGD to the point that they are mentioned in the first element of the eight original principles of OGD, "Complete — all available public data that is not subject to privacy, security or privilege limitations is made available" (Attard et al., 2015, p. 409). The other seven principles are *Primary, Timely, Accessible, Machine Processable*, *Non-Discriminatory, Non-Proprietary*, and *License-Free.* Janssen et al. (2012) included in their OGD definition "non-privacy-restricted and non-confidential data" to emphasize the importance of information privacy. Information privacy concerns are a significant barrier to expanding and improving open data initiatives in the literature (M. Janssen & van den Hoven, 2015). Barry & Bannister (2014) found that resource constraints, loss of revenue, and uncertainty concerning privacy legislation are the most prominent barriers to OGD in Ireland. Michener & Ritter (2017) identified the 'Three Ps' of resistance towards open data in the educational sector: professional, political, and personal privacy concerns. Wang et al. (2019) summarized the barriers to OGD in the literature into eight key categories, one of which was privacy restrictions. Privacy barriers in their study originated from legal compliance, public trust, and reputation concerns that might be triggered by data release (Wang et al., 2019). Overall, privacy concerns are suggested as one of the barriers/inhibitors of OGD success and value creation. These concerns are correlated with other barriers/inhibitors, such as data quality, public distrust, and unfair distribution of the costs and benefits of OGDs (Barry & Bannister, 2014; Francey & Mettler, 2021; K. Janssen & Hugelier, 2013; Wood et al., 2016).

Privacy is a significant concern in any data-sharing practice, but more so for OGDs because when data sharing is in the form of open data, minimum restrictions are imposed on users and their uses; therefore, the privacy concerns will scale up. The standard way of data release practice in OGD programs is the release-and-forget model (Ohm, 2010). The data are released to the public on the Internet, typically via an OGD portal, so the data users are unknown. There is no control over how they might use the released data. When the released data are perceived as sensitive, those concerns must be addressed first, if the sharing goals are to be realized.

Another reason for the increased importance of privacy concerns in OGD is that a large share of the data to be released as OGD by public sector bodies includes personal information about people and organizations (Van Ooijen et al., 2019). These data could be collected through voluntary channels, such as when citizens and organizations participate in surveys or censuses and provide relevant information, or through public service interaction when citizens provide their personal information in exchange for the service they receive. Moreover, enormous amounts of data are collected and stored as a byproduct of the citizens' interaction with technologies, such as mobile phones, computers, public surveillance devices, and sensors in electronic devices. These data are often recorded and stored by the Government and public institutions. Although these data were not intended for research or analytical purposes, they have a high potential for creating value if appropriately utilized. Therefore, these data could be considered potential OGD. However, an enormous amount of personal data is collected non-voluntarily through legal and regulatory tools or third parties with or without the knowledge of the data subjects.

In the context of OGD initiatives, individuals information privacy concerns have received increasing attention in scholarly articles and media outlets over the last decade. Even if the OGD initiatives, data practices are privacy-protecting and in compliance with privacy laws and regulations, the stakeholders privacy perceptions and concerns must be addressed if the governments want to reach the goals of their OGD programs. Failure to address the privacy concerns of the OGD stakeholders can create public mistrust, lack of engagement, internal protests, and, finally, prevent the programs from achieving their goals. Risks to the Government's failure to address privacy concerns can be categorized into financial, reputational, and regulatory risks (Future of Privacy Forum, 2018b).

This paper is one of the first comprehensive literature reviews about the nature of individuals information privacy concerns, the extent to which they impede the expansion and utilization of OGD, suggestions for mitigation, and how different the concerns and remedies are in different domains. Therefore, gathering, synthesizing, and organizing the extant literature is necessary to find the significant concerns, existing remedies, best practices, and theoretical and practical frameworks and identify the research gaps in these areas. The result will interest both researchers and practitioners in e-government, smart cities, privacy-preserving data mining (PPDM), privacy-preserving data publishing (PPDP), and privacy legal scholarship.

To assemble and synthesize the existing literature and provide a comprehensive report on the current state of knowledge about privacy concerns in OGD initiatives, the narrative review method was best suited from the existing methods (Templier & Paré, 2015). This narrative review explores the literature, to answer the following three research questions:

> RQ1. What are the central information privacy concerns related to the release of OGD?
> RQ2. What are the common practices and suggestions for mitigating these information privacy concerns?
> RQ3. What are the important knowledge gaps in the literature about information privacy concerns, related to the release of OGD?

Answering these questions provides decision-makers and OGD implementors to evaluate the risks and unintended consequences beforehand and optimize the benefits of the OGD programs.

Decision-makers in this context are national and local government officials. For example, city government officials can consider these concerns and risks when evaluating the appropriateness of releasing and access level of a particular dataset.

This article proceeds with a description of the research method in section 2. Next, a comprehensive discussion of the findings of this literature review is provided in section 3, which is more detailed into three subsections discussing the three research questions, including the identified privacy concerns, standard practices, proposed solutions, and remedies, and identified knowledge gaps. Section 4, discusses the findings and concludes.

## 2. Data and Methods

The narrative review method was conducted to identify, choose, and synthesize the literature on information privacy concerns in OGD programs. A narrative review is appropriate for qualitative summarization and synthesis of the extant literature and provides the readers with a state of knowledge on the topic (Grant & Booth, 2009). To report the methods and findings, this study uses the PRISMA framework guidelines (Moher et al., 2009; Page, McKenzie, et al., 2021; Page, Moher, et al., 2021) to provide a more comprehensive and reproducible report. PRISMA is a framework for reporting systematic literature review results comprehensively, and in sufficient detail to allow readers to assess the trustworthiness and applicability of the findings (Page, Moher, et al., 2021). Although this is not a systematic review, reporting the finding systematically could help the readers assess and replicate the results.

### 2.1. Search Strategy

Relevant Digital libraries were identified by surveying the literature and consulting Google Scholar. Four bibliographic databases were the most relevant: Web of Science, Science Direct, Digital ACM Library, and IEEE Explore Digital Library. These databases cover a diverse range of research areas and are exhaustive. Web of Science and Science Direct is probably the most extensive scientific collection, whereas Digital ACM Library and IEEE Explore Digital Library are perhaps the most extensive databases on digital topics. Appropriate search terms for each database were identified through forward and backward screening of the most relevant articles. Book chapters, Encyclopedia, conference abstracts, and magazines were excluded from search results. The search yielded 308 articles (190 from Web of Science, 35 from Science Direct, 27 from ACM Digital Library, and 56 from IEEE Explorer). After duplicated articles were removed from the data, 237 remained for further analysis. Table 1, shows the search queries that were used for each of the databases in January 2022.

### 2.2. Inclusion and exclusion criteria

The titles, abstracts, and keywords were reviewed for primary screening. Articles that did not discuss open government data (e.g., focused on open scientific research data, internal sharing data among government agencies, or commercial data) were excluded from the reviewed articles. Only peer-reviewed English-language articles that discussed open government data and personal information privacy concerns or privacy-preserving measures applied to OGD were included. Next, the full manuscripts were reviewed, and articles that did not discuss privacy concerns of OGD or methods for remediation of those concerns were excluded. When reviewing the full texts, relevant references to the articles were found through a manual search in Google Scholar. I consulted grey literature to strengthen our analysis. The materials consulted include but are not limited to, governmental reports and policy documents (e.g., FIPPs, privacy protection laws, such as CCPA and GDPR), reports and frameworks developed by city governments (e.g., a framework for evaluating the City of Seattle's open data program), and media reports on privacy controversies of OGD. Those that met the inclusion criteria were added to the pool of articles. Figure 1, summarizes the selection process that yielded 45 articles for inclusion.

*Table 1: Search strategy*

| Database | Search Query |
| --- | --- |
| Web of Science | (AB=("open government data" OR ogd OR "open government" OR "open data") AND AB=(privacy)) AND (DT==("ARTICLE")) |
| Science Direct | Title, abstract, keywords: privacy AND "open government data" OR "open data" OR "open government "0 |
| Digital ACM Library | [[Abstract: "open government data"] OR [Abstract: ogd] OR [Abstract: "open government"] OR [Abstract: "open data"]] AND [Abstract: privacy] |
| IEEE Explore Digital Library | "Abstract":"open government data" OR "Abstract":ogd OR "Abstract":"open data") AND ("Abstract":privacy) |

*Figure 1: The Process of Selection of Articles*

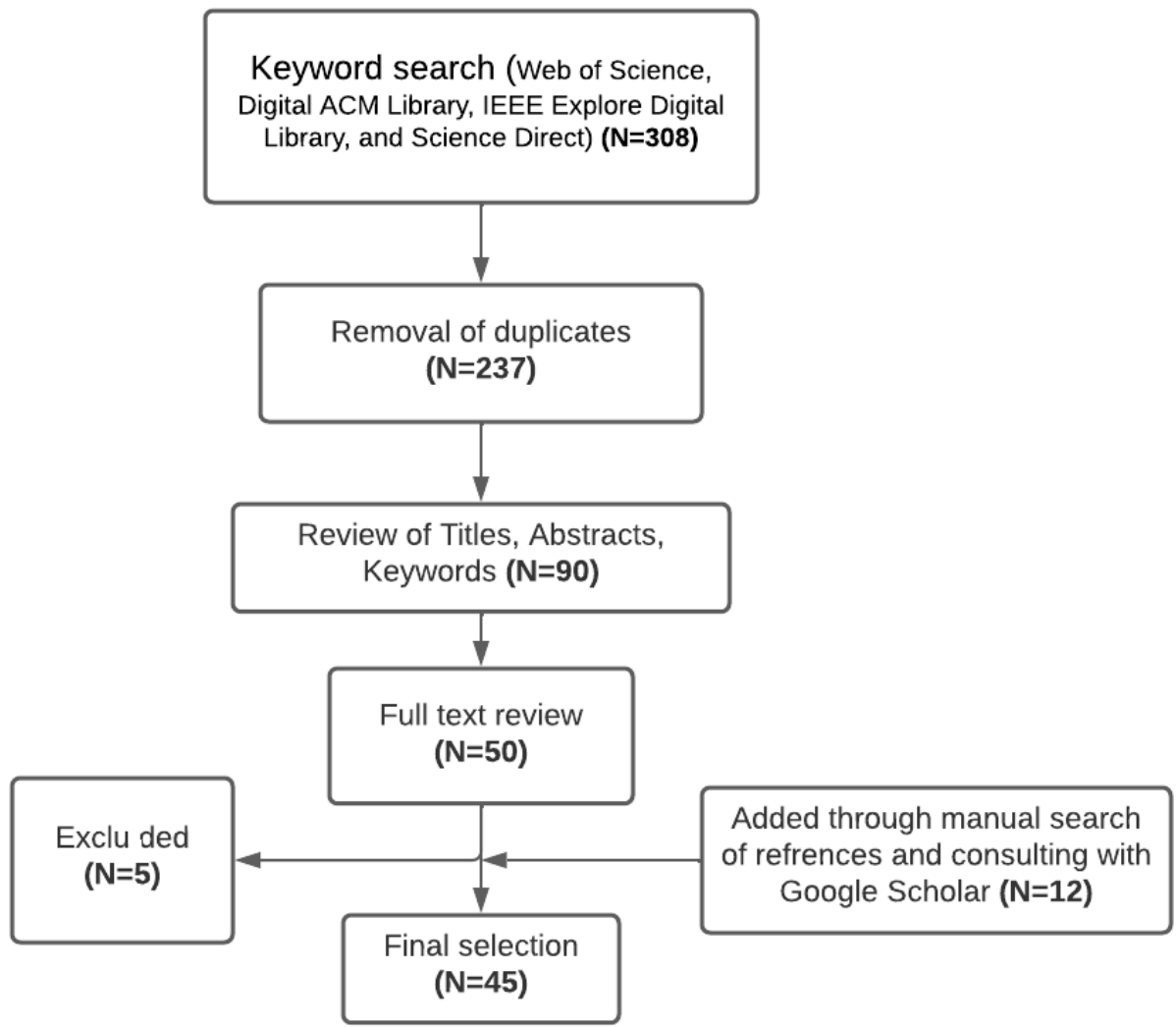

## 2.3. Review Process

A critical challenge of the synthesis was that selected articles were from heterogeneous fields of study with different research methods and reporting standards. We devised a predefined set of data elements to gather data that addressed the research questions, to overcome this challenge. In addition, selected articles' research methods, data, and research questions were collected. The data elements to be extracted were adjusted accordingly throughout the data extraction, informed by new insights from the readings. Thematic analysis was conducted to identify themes addressing the research question in the selected articles (Braun & Clarke, 2012). Section 3, reports the findings of the review.

## 2.4. Limitations

This study has several limitations. Literature on privacy issues is broad and inherently multidisciplinary. Privacy concerns of different environments are interrelated and interact with each other. The identification strategy of this study might have missed some insightful articles that have not directly discussed OGD privacy issues but those of general data-sharing practices by the public and private sectors. Some relevant articles that did not explicitly focus on OGD privacy concerns but did discuss relevant issues, were included to mitigate this limitation. The other limitation is the potential bias in academic articles that propose solutions for de-identifying OGD datasets or other technical solutions. These articles focus on risks to privacy that can be resolved. Non-academic sources were added to the pool of reviewed articles to reduce the probability of bias.

Another limitation was that only articles published in English language were considered for review. This might lead to the exclusion of some contributions from our discussion. However, most of the academic articles published in languages other than English have their abstracts in English, and those abstracts could have been captured in our search strategy, given their relevance. Therefore, we do not suspect this limitation has led to significant distortion in the findings.

## 3. Findings

All articles discussed some concerns and solutions. A minority of the articles used specific risk frameworks to elaborate the risks and benefits from the viewpoint of various stakeholders . Appendix 1, summarizes the research methods, data, and research questions of the articles, along with identified stakeholders and privacy-related barriers and challenges of OGDs. The following sections summarize the reviewed articles' identified concerns, solutions, and frameworks. Privacy concerns were about the description, analysis, criticism, and evaluation of the legal and policy environment of OGD programs concerning individuals information privacy concerns, the detrimental effects of privacy concerns on OGD goals and society, and recommendations for reforms in policymaking and legislation. The discussions about striking a balance between transparency through proactive response to a freedom of information requests and OGD initiatives, on the one hand, and individuals' right to information privacy, on the other hand, is the most prevalent topic in this stream. Solutions included technical tools, algorithmic solutions, legal and procedural measures for privacy-preserving data publishing (PPDP), including statistical disclosure limitation (SDL) techniques, de-identification algorithms and tools, tiered access designs, and privacy-preserving data mining (PPDM) techniques. Some articles have provided conceptual frameworks and practical guidelines for risk-based approaches to OGD practices for governments at different levels, including federal, state/provincial, and city/municipality.

## 3.1. Privacy Concerns

This section summarizes the literature answering RQ1, which asks what the prominent information privacy concerns are related to the release of OGD. Drawing on the reviewed literature, this section is organized into four subsections: contradictions of OGD programs with Fair Information Practice

Principles (FIPPs), reidentification risks, conflicts with OGD value propositions, and smart city privacy concerns.

### 3.1.1. Contradictions with FIPPs

Concerns about information privacy in the context of OGD can negatively affect the public sector's efficiency, citizen's rights, economic growth, and the free flow of information. For example, Borgesius et al. (2015), in their article, identified three categories of information privacy concerns regarding open data: the chilling effect on the interaction between citizens and the public sector, the lack of control over personal information, and the use of open data for social sorting and discriminatory practices. This section explores the roots of these concerns and the reasons for their importance.

The release of personally identifiable data through OGD programs contradicts some of the Fair Information Practice Principles (FIPPs). FIPPs were initially introduced in 1970 by the U.S. Department of Health, Education & Welfare, in response to growing privacy concerns over storing and processing information about an individual with computer systems. Over time, the principles constantly evolved by various organizations and have formed the foundations of some prominent aspects of modern information privacy laws around the world, including the European 2018 General Data Protection Regulations (GDPR) and 2020 California Consumer Privacy Act (CCPA), and framework for privacy policies and standards in private and public organizations (Gellman, 2021). In what follows, we review the most internationally acknowledged version of the principles, developed by the Organization for Economic Cooperation and Development (OECD, 2013) and discuss their extant conflicts with common OGD practices, extracted from the reviewed articles. Table 2, lists the eight FIPPs along with a description extracted from the OECD privacy framework (OECD, 2013). The OECD framework applies to personal data, defined as "any information relating to an identified or identifiable individual."

*Table 2: FIPPs proposed by Organization for Economic Cooperation and Development*

| Principle Title | Principle Description |
|---|---|
| Collection Limitation Principle. | "There should be limits to the collection of personal data, and any such data should be obtained by lawful and fair means and, where appropriate, with the knowledge or consent of the data subject." (OECD, 2013) |
| Data Quality Principle. | "Personal data should be relevant to the purposes for which they are to be used and, to the extent necessary for those purposes, should be accurate, complete, and kept up to date." (OECD, 2013) |
| Purpose Specification Principle. | "The purposes for which personal data are collected, should be specified no later than the time of data collection, and the subsequent use, limited to the fulfilment of those purposes or such others as are not incompatible with those purposes and as are specified on each occasion of change of purpose." (OECD, 2013) |

| Use Limitation Principle. | "Personal data should not be disclosed, made available or otherwise used for purposes other than those specified, except a) with the consent of the data subject, or b) by the authority of law." (OECD, 2013) |
|---|---|
| Security Safeguards Principle. | "Personal data should be protected by reasonable security safeguards against such risks as loss or unauthorized access, destruction, use, modification or disclosure of data." (OECD, 2013) |
| Openness Principle. | "There should be a general policy of openness about developments, practices, and policies, concerning personal data. Means should be readily available of establishing the existence and nature of personal data and the main purposes of their use, as well as the identity and usual residence of the data controller." (OECD, 2013) |
| Individual Participation Principle. | "An individual should have the right: a) to obtain from a data controller, or otherwise, confirmation of whether or not the data controller has data relating to him; b) to have data relating to him communicated to him, within a reasonable time, at a charge, if any, that is not excessive; in a reasonable manner, and in a form that is readily intelligible to him; c) to be given reasons if a request made under subparagraphs (a) and (b) is denied and to be able to challenge such denial; and d) to challenge data relating to him and, if the challenge is successful, to have the data erased, rectified, completed or amended." (OECD, 2013) |
| The Accountability Principle. | "A data controller should be accountable for complying with measures which give effect to the principles stated above." (OECD, 2013) |

Typically, OGD are collected through the interaction between citizens and government agencies in the service provision process as a byproduct. In this scenario, the consent of the data subject is irrelevant, and there is not much choice for them. However, cases of excessive data collection that violate the collection limitation principle are frequently reported in the literature (Huang et al., 2019; Wood et al., 2016). Data collection practices of national governments are more restricted, due to regulations, than private companies. On the other hand, local governments at the municipality level create more privacy concerns and controversies with their excessive data collection without individual consent. For example, in addition to data from city registers, a large amount of data from traffic control cameras, public transportation usage, air quality monitors, facial recognition devices, security cameras, and automatic plate number readers are collected without the knowledge of the data subject (Rohunen et al., 2014; Rubinstein, 2018; van Zoonen, 2016). World Bank's Policy Research Working Paper 9811 (Jolliffe et al., 2021) identifies 12 conditions for data created by the public sector to be valuable and have a transformational impact. The conditions are grouped into four categories: spatial and temporal coverage (complete, frequent, and timely), high quality (accurate, comparable, and granular), easy to use (are accessible, understandable, and interoperable), and safe to use (are impartial, confidential, and appropriate). The first three groups of the conditions are addressed in the literature on OGD utilization and barriers (Safarov et al., 2017). The last group of conditions is the centre of privacy concerns in the reviewed articles in this study. Impartiality of OGD reduces the risk of discriminatory effects of OGD release, usage, and analysis on individuals and population groups (Borgesius et al., 2015; Green et al., 2017). However, ensuring the impartiality of OGD is challenging, when data are administrative or collected indiscriminately through surveillance programs, sensor devices, and public engagement initiatives. Preserving the confidentiality of OGD is usually practiced through de-identification methods, which will be discussed in the next subsection. A common challenge in de-identification efforts identified in the literature is the adverse effects of de-identification on released data quality in terms of other data quality dimensions, including coverage, accuracy, and impartiality (Hasanzadeh et al., 2020; Heijlen & Crompvoets, 2021; Rantala et al., 2020).

The principal assumption behind the innovation and economic value of OGD is that the released data are not only visited on OGD portals but also reused by developers, journalists, and public and private companies to create social and economic value (Safarov et al., 2017). Therefore, specification of purpose at the time of data collection for OGD is almost impossible given the unknown users and use cases and open license requirement of OGD (Kjærgaard et al., 2020).

OGD data are being shared publicly for purposes other than those specified at the time of collection. Although some suggested notice and consent mechanisms for OGD, like what has long been practiced in scientific research data practices (Altman et al., 2018; Vasileva et al., 2018), obtaining informed consent in the OGD context appears more problematic. Governments encourage the reuse and redistribution of OGD as an essential dimension of OGD goals. The other issue arises from the intrinsic power asymmetry in interactions between a government agency and individual data subjects. The lack of individual control over personal information is seen as more problematic in public-sector interactions than in the private sector (Borgesius et al., 2015).

Protection of personal data from disclosure and unauthorized access, destruction, use, and modification in the context of OGD is either relevant in the data lifecycle's collection, storage, and retention stages or the data release stage. In the release stage of OGD, privacy protection and security safeguards are often considered the same and used interchangeably. Examples of security breaches in the pre-release stages are unauthorized access to raw data by third parties and data storage in an insecure cloud environment (Moustaka et al., 2019; Rubinstein, 2018). Henriksen-Bulmer et al. (2019) Observe that in the OGD community, privacy risk is often considered one element of security risk, undermining privacy risk assessments.

Openness and transparency about personal data processing are the most important principles in data protection laws such as GDPR. In the context of OGD, data subjects should be informed about whether and what information about them is released or used to make summary statistics and visualizations and what privacy-preserving techniques, legal, procedural, and organizational measures are implemented to protect them from privacy violations and related future harms (Future of Privacy Forum, 2018a; Henriksen-Bulmer et al., 2019).

Data subjects' rights to collect and modify personal data can be supported within the existing government, legal and organizational frameworks. Some governments, including national and federal levels, and recently some states have enacted laws that guarantee the data subjects' rights to know if and what data about them are retained and modify the data if necessary (for example chapter 3 of GDPR (REGULATION (E.U.) 2016/679 OF THE EUROPEAN PARLIAMENT AND OF THE COUNCIL, 2016)). Some laws provide the same rights for sector-specific sensitive data, such as educational and financial data (Ohm, 2015). However, in the context of OGD, we could not find any practical cases of engaging data subjects in their personal data processing and release.

The accountability principle is closely related to all other principles of fair information practices. Specifically, it can be improved by ensuring openness and individual participation principles. However, legal and litigation procedures are required to ensure accountability principles. In the OGD context, the controller is the government agency that decides whether and with what level of accuracy and granularity the personal data can be released through OGD programs. The high degree of accountability and legal and social liability could improve the Government's precautious acts in the process of sanitizing and releasing OGD (Future of Privacy Forum, 2018a; Michener & Ritter, 2017).

Some articles in the literature suggest improving consent mechanisms in personal data collection practices, by government bodies, to mitigate the privacy concerns of OGD programs raised from contradictions with FIPPs. For example, (Vasileva et al., 2018) suggest that establishing a consent process for data collection (and opt-out option) and communicating data usage and data protection procedures could increase the stakeholders' trust. However, we could not find any practical case of obtaining individual consent for OGD programs. In their proposed conceptualizing framework for transparency and privacy in the context of OGD, (M. Janssen & van den Hoven, 2015) identified the consent mechanism in existing privacy frameworks in European countries, as the main barrier to the utilization of linked OGD.

Literature on the inadequacy of the notice and consent approach of the private sector for privacy protection is rich (Kroger et al., 2021; WEF, 2020). Data subjects can hardly observe the collection, use, and transfer of their personal data when interacting with public services, even if policy disclosures follow the guidelines thoroughly. Data subjects cannot comprehend the full impacts and consequences of collecting and disseminating personal information. Bounded rationality and cognitive biases have been shown to impose severe limitations on people's decision-making concerning their privacy (Acquisti et al., 2016). The fundamental assumption of OGD is that everyone can use public data without limitation on usage or reuse and re-sharing. This assumption contrasts with the concept of notice and consent, which is supposed to clarify for the data subject what purposes the data will be used.

### 3.1.2. Reidentification risks

The FIPPs, laws, and regulations derived from them only apply to Personally Identifiable Information (PII). Even in a modern data protection law like GDPR, anonymized data are no longer considered personal data and lose the protections devised in the regulation. However, progress in the algorithms and infrastructure of data analysis and the extraction of information from big data in the last decade has shown us that, complete anonymization can be achieved only by not releasing any data. Deidentified data published through OGD portals can be re-identified with the help of external auxiliary information or through record linkage to other available datasets.

Future of Privacy Forum (FPF) has identified three core privacy risks relating to open data programs of local governments: reidentification risk, biased or inaccurate data, and loss of public trust (chilling effect). Although Ohm's (2010) prediction of privacy experts losing faith in anonymization has not yet come entirely true, there are increasing concerns about the ease of reidentification of individuals from publicly released databases.

Some characteristics of OGD distinguish them in terms of reidentification risks and corresponding precautions to avoid them. First, even if anonymized, the published personal data could be accessed, analysed, de-anonymized, and manipulated by any party with different capabilities. Releasing de-identified personal information as open data could increase the risk of re-identifying individuals, groups, and organizations through the reuse of released data combined with other private and public datasets by third parties (Future of Privacy Forum, 2018b; Huang et al., 2019; K. Janssen & Hugelier, 2013; Kjærgaard et al., 2020; Zhao, 2008). The publisher could not possibly know what additional information other possible parties have that might help them re-identify the published OGD (Finck & Pallas, 2020). Popular initiatives, such as smart cities, smart universities, smart transportation, and the like, promote an environment in which all activities and behaviours of individuals are recorded, analysed, combined with other data, and used for profiling (Scassa, 2014).

The second concern is that when OGD is released, it cannot be retracted anymore. The data could be downloaded and recirculated by other unknowable parties outside the publisher's control. The unpredictability of technological improvement in data analytics and hardware adds to this complication. Specifically, suppose personal data with lasting relevance (e.g., criminal history of individuals) gets published in the form of OGD. In that case, the detrimental consequences to the data subjects stay long even if the responsible agency notices the personal information leak and removes the data from its portal.

Even if adequately de-identified datasets released through open data programs might pose a serious privacy risk to the data subjects, because they could be a valuable source of data for potential intruders (Austin & Lie, 2019; Gkoulalas-Divanis & Mac Aonghusa, 2014a; Huang et al., 2019; Lavrenovs & Podins, 2016). Other datasets published with weak de-identification by other public organizations or commercial entities and personal data shared through social media platforms and other sources, can significantly increase the risk of reidentification.

### 3.1.3. Privacy vs. OGD Value Propositions

Creating a balance between individuals information privacy rights and OGD value propositions, including transparency and accountability, participation and collaboration, innovation and economic growth, and government efficiency and effectiveness, is one of the recurring themes in the reviewed literature. Meijer et al. (2014) Suggest that interpreting OGD initiatives in the public value paradigm might help understand OGD policy problems and resolve the contradictions between public values. They identified transparency and trust as public values contrasting with two other values of privacy and security.

Privacy and Transparency/accountability trade-off has gained more attention, specifically in law review articles. This balance could be seen and thought of as a particular case of the balance between privacy and the free flow of information (Ohm, 2010). The Government's release of public personal information through OGD portals is perceived to enhance transparency and accountability of the public sector in general, through informing the public and increasing their engagement and participation.

In their highly debated work, (Bannister & Connolly, 2011) argue that transparency, specifically e-transparency, is not in itself and always a good outcome for good Government and governance. One of the primary reasons they make their arguments is the threat the increased transparency imposes on the privacy of citizens and public servants. They argue in favour of weighing privacy and other risks against the benefits of transparency. In another contribution, (K. Janssen & Hugelier, 2013) explores the balance between Freedom Of Information (FOI) acts and information privacy and data protection laws to find criteria and balance processes to apply to personal data in an open data environment. They argue that the proactive approach of open data, rather than the reactive/responsive approach of FOI, and the economic and innovation purpose of open data, as opposed to accountability and transparency-focused objectives of FOI, make the dynamics and mechanisms of a balance in the two subjects different.

A clear distinction between the transparency/accountability goals of OGD and its innovation/economic goals could help understand the above balance. OGD decision-makers need to know which datasets and compliance with open data standards can promote transparency/accountability or innovation/economic goals. (Conroy & Scassa, 2015) observe that the trade-off is usually between privacy and economic value, rather than privacy and transparency. Privacy concerns and suggested solutions identified in smart cities literature (van Zoonen, 2016) approve this observation. In the next section, we briefly discuss the finding on privacy concerns of smart cities in the reviewed articles.

Finally, the open Government's transparency/accountability and public participation goals can be achieved without releasing raw data containing personal information. Based on the analysis of five case studies in the U.S., Lee & Kwak (2012) developed an open Government maturity model for social media pubic engagement, starting from initial conditions at the lowest maturity level and ending with ubiquitous engagement at the highest level.

### 3.1.4. Privacy Concerns in Smart Cities

The emphasis on openness (open data standards and open licensing), granularity, timeliness, accessibility, usability, and single-portal release of OGD in existing policies indicate the dominance of innovation and economic value, rather than other goals, such as transparency, accountability, and participation which could be addressed through other aspects of open Government initiatives, including open access and open engagement (Scassa, 2014). The innovation and economic growth aspects of OGD programs are more evident in smart cities literature (Khan et al., 2017; Piao et al., 2019; van Zoonen, 2016).

As more services are provided to citizens through a network of connected technologies and new services and innovations are invented daily, cities' amount and complexity of data generated and collected are striking. City data are diverse and differ in granularity, frequency, timeliness, and complexity. This variation results from diverse sources of data and a broad range of technologies used to collect and combine the data. For example, (van Zoonen, 2016) provides a city data landscape that identifies six-city sectors as creators and collectors of city data. The sectors include infrastructure (transport and asset management, built environment), sustainability (Energy usage, water, environment, weather), health (Health, quality of life, well-being, life expectancy), cohesion (Education, social capital, migration, neighbourhoods, housing, crime), commerce (Business opportunities, marketing, location-based services), experience (Events, leisure, nightlife, tourism, heritage). Managing the privacy concerns and risks of OGD programs at the municipal level becomes challenging when considering this complexity.

Van Zoonen (2016) provides a framework for understanding and managing data collection and user privacy concerns in smart cities. In a two-by-two scheme, data is categorized based on its sensitivity and the purpose of collection. Sensitivity refers to whether the collected data are personal or not. The purpose dimension has two ends: service and surveillance. Intuitively, the author suggests that personal data collected for surveillance raises the most privacy concerns and should be treated more carefully.

Managing privacy concerns of cities' OGD programs has more nuances. First, impersonal data collected from traffic flow, public transport, CCTV, utility sensors, and environmental sensors, when released to the public through OGD portals and reused vastly by researchers, developers, and data analysts, could become personal (Gao & Janssen, 2020; Khan et al., 2017; J. S. Lee & Jun, 2021; Shaham et al., 2021; van Zoonen, 2016; Vasileva et al., 2018). Second, the benefits arising from the smart city data release seem more diverse and can serve different goals. Published city data could be utilized for diverse goals, from social justice advocacy to academic research and civic engagement to commercial uses by companies (Rantala et al., 2020; Sinnott et al., 2016) and individual developers (Ansari et al., 2022; Hasanzadeh et al., 2020).

### 3.2. Solutions for Mitigating Privacy Concerns

This section introduces the findings from the reviewed articles to RQ2: What are the standard practices and suggestions for mitigating privacy concerns? Proposed solutions for mitigating the privacy concerns originating from OGD initiatives include technical, legal, and procedural instruments. This section reports the suggested solutions we find in the reviewed articles. Table 3, summarizes the solutions along with a short description of them.

*Table 3: Solutions for mitigating OGD privacy concerns*

| Solution | Description |
| --- | --- |
| Technical | Technical solutions comprise methods and techniques for de-identifying personal data and measures for identifying and |

|  | lowering the risk of reidentification of individuals from the released datasets. |
|---|---|
| Legal and Procedural | Relying on a combination of procedural, educational, economic, and legal controls for privacy protection. These measures include privacy-by-design, privacy impact assessment, internal review boards, restriction and ban of reidentification. |
| Risk-based Frameworks | This group lends from information security risk management and offers frameworks for addressing OGD privacy concerns through assessing the risks and devising mitigation strategies. |

### 3.2.1. Technical Solutions

#### 3.2.1.1. Privacy Models

Before reviewing the de-identification techniques, two of the most recurrent modern privacy models in the reviewed literature are presented below: k-anonymity and differential privacy.

K-anonymity and extensions

A well-known formal privacy model for protecting a dataset from reidentification is k-anonymity. The primary goal of the k-anonymity mechanism is to guarantee that if an attacker tries to link an individual to a record based on its quasi-identifiers, there exist at least k records in the dataset that cannot be distinguished (Samarati & Sweeney, 1998). Each group of these indistinguishable records makes an equivalence class. K-anonymity is usually achieved through the generalization and suppression of records. The concept of minimal generalization is core to k-anonymity, and it states that the amount of distortion imposed on the data must not be more than needed. Therefore, it implicitly considers a tradeoff between data utility and marginal privacy protection. Later evaluations, however, show that k-anonymity methods usually over-anonymize data resulting in excessive distortions to the data and information loss (El Emam & Dankar, 2008).

Another privacy method is l-diversity which is proposed to overcome the limitations of the k-anonymity method. Even with a large k and equivalence classes, a risk of attribute disclosure still exists. This risk originates from the low diversity of the targeted sensitive attribute. For example, in a 10-anonymized dataset, there are 10 indistinguishable records regarding all quasi-identifier combinations. If the sensitive attribute has a low diversity of only two values, including 8, "disease A" and 2, "disease B," then one can infer that an individual belonging to that class has disease A with a probability of 80 percent. The l-diversity method requires at least l well-represented value in each equivalent class. In the above example, then, in order to have 5-diversity anonymity, there should be at least five well-represented values of diseases in the sensitive attribute. This method also protects against background knowledge attacks, where k-anonymity is vulnerable (El Emam & Dankar, 2008; Sweeney, 2002). Li et al. (2007) show that combining k-anonymity and l-diversity requirements does not protect data from attribute disclosure. They proposed the notion of t-closeness, which requires that the distribution of the values of sensitive attributes in each and any equivalence class must be close to the distribution of the attribute in the whole dataset.

Differential Privacy

The computer science approach to privacy protection has sought to offer a formal guarantee of privacy to a given dataset without assumptions about the knowledge of the attacker and attack models. The most well-known and practiced formal privacy model was introduced by Dwork (2006) and has been applied in various fields of study, from health data (Dankar & El Emam, 2012) to census microdata publication (Abowd et al., 2020).

Although differential privacy offers a formal measure of privacy loss, with rigorous mathematical guarantees, it has limitations and weaknesses. Differential privacy cannot preserve the full utility of the data, while providing a level of privacy protection, even if it is more efficient than other methods. Most research for developing differential privacy algorithms, concerns interactive query settings that are not compatible with OGD practices, where the complete dataset is published publicly, and the data users have no limitation in reusing the data.

The differential privacy model has been employed by tech corporations like Google for years and has recently been employed by some government entities like Census Bureau (Abowd et al., 2020). However, limitations and unintended consequences of performing differential privacy on datasets have received attention in the literature. Differential privacy machine learning models are shown to worsen disparity in model accuracy, increasing the false classification of underrepresented groups (Bagdasaryan et al., 2019). Differential private versions of traditional statistical datasets make substantial distortions in the data making social and economic inequalities larger (see, for example, Hauer & Santos-Lozada (2021)).

### 3.2.1.2.   De-identification Techniques

Agencies take some privacy protection measures before releasing open data, including de-identifying sensitive personal data, aggregating data and releasing summary statistics, and data anonymization. This section provides an overview of common privacy-preserving techniques used in industry and government agencies, which include suppression, generalization, and synthetic data.

Suppression

Suppression of the entire attribute or suppression at the record level is the simplest way to de-identify a given dataset. For example, the name field in the dataset is a direct identifier and might be omitted entirely before sharing. Similarly, some records with high identifiability, for example, outliers in income attributes, may be dropped from the database. Although simple to implement, suppression incurs a high data utility loss and might even make the dataset completely useless (Kjaergaard et al., 2020; Matthews & Harel, 2011).

Generalization

One main non-perturbative disclosure control method is data generalization, when numerical values are replaced with intervals or rounded to discrete values. This method has been a common tool for statistical agencies in preserving the privacy of individuals but is not effective in many cases (Groves & Harris-Kojetin, 2018).

Synthetic data

A different perturbative approach for protecting the privacy of the subjects of microdata is to generate synthetic data from original data and publish it instead of the original data. The initial idea was proposed by Rubin (1993) as an alternative to masking micro-data to protect the confidentiality of the data subjects. Recently there has been a growing interest in publishing public sector microdata in synthetic datasets (Steinbakk et al., 2020). A well-known example of synthetic data published by government agencies for public use is SIPP Synthetic Beta (SSB), a synthetic integration of person-level micro-data from a household survey with administrative tax and benefit data (U.S. Census Bureau, 2015).

To be effective and useful, synthetic data must satisfy two properties. First, it must preserve the underlying patterns of the original dataset or, in other words, the joint probability distribution of the variables as much as possible. Second, the risk of identity or attribute disclosure should be minimized. Therefore, a spectrum can be imagined in which the more the synthesized dataset mimics the underlying data-generating process of original data, the more risk of disclosure exists. Given that the data-generating process is seldom known to the analysts or data owners, it is practically impossible to create synthetic data that completely preserves the original data-generating process.

The quality of synthetic data should be assessed in two dimensions. Utility measures quantify the analytical value of synthetic data, in terms of how close the inferences made based on synthetic data, are to that of the original dataset. Some scholars suggest that first, the validity of the synthetic data in terms of the format, data types, and the plausibility of the combinations of variables be established, and then assess the similarity of inferences between the two datasets (Steinbakk et al., 2020). Utility measures are also grouped into two types of general and specific measures. The former assesses the similarity between the synthetic and original data in terms of, for example, the distance between the distributions of the variables or propensity score measures.

On the other hand, the specific utility compares the outputs of specific models employed on the two datasets (e.g., the regression coefficients and their confidence intervals). Visual examination of the similarity between the synthetic and original data is common in literature. Comparisons of the distributions of the variables with histograms/boxplots or the comparison of the correlations among the variables by scatterplots are examples of these types of analyses. Given the potential of synthetic data to reduce the disclosure risk to the desired level, open government programs could naturally be a major use case for synthetic data. Calcraft et al. (2021) identified three use cases for synthetic data in government and public policy research: exploratory analysis without access to original data, training researchers on how to handle challenging datasets, and writing and testing analysis code before getting access to original data. They classified synthetic data into two groups of low-fidelity and high-fidelity when the former indicates that the synthetic data only preserves the format and structure of the original data, and the latter refers to data that mimics the underlying patterns of the original data, including the relationships among the variables. Although, in general, there are fewer privacy concerns in publishing synthetic data, high-fidelity synthetic data may still have a considerable risk of disclosure. Other privacy-preserving techniques should be accompanied when necessary to reduce the risks, especially when the intent is to open up the dataset to the public.

In an open data context, most of the implementations of synthetic data have been in health data. (Quintana, 2020) provides a guide for using the R package, synthpop, to generate synthetic biobehavioral datasets to share in open data repositories. Using more advanced A.I. techniques, (Yoon et al., 2020) implemented Generative Adversarial Networks (GANs) to generate synthetic data that closely approximates the joint probability distribution of the variable in the original Electronic Health Records (EHR) dataset. (Yazdanie & Orehounig, 2021) suggest using synthetic data for privacy preservation in the context of energy consumption open data.

### 3.2.2. Legal and Procedural Solutions

Going beyond technical solutions, another theme in the literature explores different legal and procedural solutions for mitigating the privacy risks of OGD programs. Altman et al. (2018) suggest relying on a combination of procedural, educational, economic, and legal controls for privacy protection rather than simple interventions like de-identification or consent. Tiered access to government data, conducting privacy risk assessment (with consideration of data frequency, dimensionality, age, and sensitivity), and internal review boards are some interventions suggested by Altman et al. (2018). Some of these suggestions, such as tiered access and notice and consent, are provided by building on traditional privacy practices in long-term research studies. Therefore, the applicability of these legal and procedural controls in the context of OGD with explicit requirements of free public access to the data and the nature of public data as a byproduct of government services is ambiguous. Other procedural controls, such as risk assessment (Borgesius et al., 2015; Green et al., 2017; Michener & Ritter, 2017; Scassa & Conroy, 2017; Wang et al., 2019) and internal review boards (B. Lee et al., 2021) seem promising in managing privacy risks of OGD programs. Section 3.2.3 explores the risk-based approaches in more depth.

Banning data users from reidentification of OGD datasets, access restriction, and giving individuals control over their data have been proposed in some articles (M. Janssen & van den Hoven, 2015; Kjaergaard et al., 2020; Meijer et al., 2014; Rohunen et al., 2014; Scassa & Conroy, 2017; van Loenen et al., 2016; Vasileva et al., 2018; Walsh et al., 2018). However, there are serious doubts about the practicality and effectiveness of this approach. Both discovery and enforcement of the ban and restriction could be challenging to the diverse and unknown population of OGD consumers.

### 3.2.3. Risk-based Frameworks

Several articles have discussed risk-based frameworks. Three main points were mentioned in these articles: considering a spectrum of openness for OGDs, OGD risk assessments, and OGD evaluation frameworks. The current definitions of open data imply that there should not be any restriction on public sector data access and use/reuse. Implementing a more moderate spectrum-based approach could increase the amount and utilization of the released OGD without endangering personal information privacy. Releasing public sector data with restrictions on the use and access may not be compatible with current open data principles. Still, it could help achieve the open data goals with negligible privacy loss.

A common theme in the literature originates from a risk and outcome-based approach to managing information challenges in organizations. The idea behind this approach is that the risk of personal information disclosure and violation of individual privacy cannot be eliminated without giving up the benefits of releasing public data. A risk-based approach to information policy is well-known in the information security community but less in privacy-related practices.

Version 1.0 of the NIST Privacy Framework was published in January 2020 by the U.S. national institute of Standards and Technology. The privacy framework aims to help organizations manage privacy risks related to their information processing practices while deriving maximum benefits from their data processing (National Institute of Standards and Technology, 2020a). The framework follows the structure of the NIST cybersecurity framework to facilitate the integration of privacy and security risks into the enterprise risk management framework. Similar to the cybersecurity framework, the privacy framework consists of the core, profiles, and implementation tiers. A granular set of activities and outcomes for enabling dialogue about managing privacy is collected in the core section. The core consists of functions, categories, and subcategories. A selection of specific functions, categories, and subcategories from the core comprises a profile. These profiles can be used to describe the current state, or desired target state, of the organization in terms of privacy risk management activities and outcomes. The difference between the two profiles could help the organization identify the gaps, develop an action plan for improvement, and estimate the resources it would need to achieve the privacy outcomes. These profiles can also communicate the organization's privacy risks with external entities, including the public. The framework is designed with flexibility in mind to enable different types of organizations to adapt and employ it based on their "mission or business objectives, privacy values, and risk tolerance; role(s) in the data processing ecosystem or industry sector; legal/regulatory requirements and industry best practices; risk management priorities and resources; and the privacy needs of individuals who are directly or indirectly served or affected by an organization's systems, products, or services." (National Institute of Standards and Technology, 2020a). Government entities at different levels could develop target and current profiles based on their resources, strategic plans, legal and regulatory environment, and citizens privacy concerns. This is specifically true for local governments at the city level with their own economic and operational needs, regulatory specifications, and public concerns and is not bound by privacy rules that apply to federal agencies. Implementation tires are incorporated in the framework to support organizations in selecting proper target profiles while considering their resources and needs. Four distinct tires, partial, risk-informed, repeatable, and adaptive, are defined in the framework, representing a progressive path of privacy risk management practices.

Although the NIST Privacy framework is designed to be flexible and usable for different organizational settings, the specific context and peculiarities of OGD initiatives might hamper the usability and efficiency of the framework to manage privacy risks of OGD practices. However, the framework could be used as a scheme to design and develop OGD-specific risk frameworks and procedures, by local and national governments, to address their specific legal, economic, and participatory needs. The notions of transparency, accountability, citizen participation, and other potential outcomes of the OGD initiative, along with potential privacy risks, such as a chilling effect on public participation, discriminatory effects on subpopulation, and worsening inequality must be added to the framework when evaluating the benefits and risks of releasing a dataset.

NIST Special Publication 800-53 Revision 5, published in September 2020, provides a comprehensive list of Security and Privacy Controls for Information Systems and Organizations. The controls in the publication are flexible and customizable for different types of organizations with different needs and resources. The controls address security and privacy from functionality and assurance perspectives and can be implemented within any organization or system that processes, stores, or transmits information. Federal government agencies must implement and comply with the controls in the publication, but it "is designed to help organizations identify the security and privacy controls needed to manage risk and satisfy the security and privacy requirements" (National Institute of Standards and Technology, 2020b). Of the 20 control families in NIST SP 800-53, two are specifically relevant to OGD policies and practices: PII Processing and Transparency and Risk Assessment.

Henriksen-Bulmer et al. (2019) provide a meta-model for open data decision-making, based on Nissenbaum's (2009) contextual integrity framework that can potentially be incorporated into a privacy risk assessment and decision-making support tool. Their model includes three phases of explanation, risk assessment, and decision built on the "3 Key Elements" in Nissenbaum's language. Nine decision heuristics are used to navigate the decision-making process by open data managers. In the explanation phase, mapping and assessment of the current information flow system are conducted by identifying four decision heuristics: dataset, actors and roles, context, and information transmission practices. The risk assessment phase evaluates privacy risks to identify how any change in the current information flow would pose a risk to privacy. Norms, values, regulations, and disclosure risks are considered in the risk assessment phase. Based on the findings of the first and second phases, the decision is made about whether a change in information flow is compatible with contextual integrity or not. The contextual integrity model for open data was applied to a local authority in the U.K., to assess the privacy risks of three datasets that had already been published and found that two of the datasets were deemed unsuitable for publication.

Each dataset has its risk-benefit trade-off. Estimating the expected benefits and costs is complicated and unpredictable. Risk assessment before releasing any dataset could be costly for public sector agencies and, as a result, prove counterproductive. Evaluation of government open data programs regarding privacy, security, and data practices considerations can help the public and governments learn the strengths and weaknesses and modify and improve the procedures and processes. For example, (Future of Privacy Forum, 2018a) has developed a framework for evaluating the City of Seattle's open data program in six domains based on the AICPA/CICA Privacy Maturity Model (PMM). The six domains included privacy leadership and program management, benefit-risk assessment, de-identification tools and strategies, data quality, equity and fairness, and transparency and public engagement.

## 3.3. Knowledge Gaps

This section introduces the findings to the RQ3 from the reviewed articles: What are the important knowledge gaps and future research directions? Several knowledge gaps were identified through this literature review. This section discusses these gaps and open questions for further inquiry. First, a lack of empirical evaluations of the extent to which the existing opened datasets contribute to privacy threats to individuals is evident from the existing literature. Many of the privacy concerns and risks discussed in the literature refer to potential hypothetical scenarios, such as those listed in Appendix 1. Although these scenarios are likely to happen, a need for more empirical solid evidence for the actual prevalence and magnitude of those risks can lead to more robust and practical discussions. One primary source of this lack of empirical evidence is that actual uses of OGD are challenging to track. Unlike the vast body of literature on the supply side, there is a lack of research in identifying and understanding the demand side of the OGD ecosystem (Dawes et al., 2016). A deep understanding of the actual use cases of OGD can guide policymakers in selecting proper datasets to open to the public with minimum risks and maximum benefits.

Given the public open access to the OGD datasets, users of these datasets are primarily unknown (Gkoulalas-Divanis & Mac Aonghusa, 2014a). Studies that can identify those users who exploit OGD datasets, to identify individuals for profiling and advertising purposes, are needed. Knowing those groups of users, their incentives, and use cases could inform decisions to limit the privacy threats without losing the benefits of OGD. For example, Federal Trade Commission released a report in 2014 exploring the practices of data brokers using data from different sources to make individual profiles for marketing, risk mitigation products, such as identity verification and fraud detection, and people search products (Federal Trade Commission, 2014). In a more recent study, building on Davies's (2010) ontology, (Magalhaes & Roseira, 2020) studied the commercial uses of OGD data in 178, for-profit organizations in the United States. It found further support for the suggested five themes. However, this study is based on self-reported accounts of the OGD user companies. It does not make a connection between specific OGD datasets and use cases and thus, barely adds to our understanding of the actual risks and benefits of OGD.

The other understudied topic in the literature is the identification of the distributional impacts of OGD initiatives concerning privacy threats. Although many use cases of civil society activists, journalists, NGOs, and civic app developers are documented in the literature (Chen et al., 2017; Henninger, 2018; M. Janssen & van den Hoven, 2015; van Zoonen, 2016), the lack of robust empirical evidence of the distribution of the benefits and risks of OGD programs is still evident. More specifically, policymakers, risk evaluators, and the public would want to know the demographic distribution of those who benefit and those who bear the costs of OGD, including privacy costs.

As discussed in section 3.3, a recurrent theme in the literature calls for embracing risk-based approaches in OGD decision-making. Some extant risk frameworks could guide government entities' OGD program design and implementation (National Institute of Standards and Technology, 2020a). However, given the specific context of OGD with the complexities of identifying the risks and find-ing mitigations, there is a need for specialized risk framework OGD programs. These risk frameworks could be designed for different levels of Government with different needs and complexities. Henriksen-Bulmer et al. (2019) provide an excellent example of such efforts to be done.

## 4. Discussion and Future Research Directions

Open government data initiatives have gained popularity among different levels of Government during the last decade. National and state governments, sectoral government agencies, and regulatory organizations have created online portals and released different data to enhance transparency, accountability, innovation and economic growth, efficiency, and effectiveness of government functions. At the city level, open data initiatives serve as a major component of the smart city paradigm with more attention to innovation, economic boost, and service provision than other public values. Public data released through OGD initiatives might contain personal data or could be combined with other sources of private and public data to gain personal information about an individual. Disclosure of personal information about people exposes them to a myriad of harms and risks, including but not limited to identity theft, financial scams, reputational damage, lack of employability, lack of insurability, discrimination, and surveillance (Barati & Ansari, 2022; Barati & Yankson, 2022; Francey & Mettler, 2021; Institute of Medicine, 2013; Pernelle et al., 2013; Wood et al., 2016). Specific harms can arise from personal privacy violations of OGD programs, including loss of public trust in government institutions, chilling effect on free expression, and unintentional disclosure of national security information. This study reviewed the state of the knowledge regarding prominent privacy concerns and risks from OGD programs in all levels of Government and public sectors and the technological, organizational, and legal solutions and frameworks proposed in the literature for managing those concerns and risks.

The first set of privacy concerns identified in the reviewed articles were contradictions with fair information practices. Specifically, intrinsic conflict of OGD goals and practices with purpose specification and use limitation principles have been discussed in the reviewed articles (Gkoulalas-Divanis & Mac Aonghusa, 2014b; Kjaergaard et al., 2020; Meijer et al., 2014). Recommendations are proposed for notice and consent mechanisms in data collection and release processes, for example, by Altman et al. (2018), Vasileva et al. (2018), and Janssen & van den Hoven (2015). However, the practicality and effectiveness of those mechanisms have yet to be empirically evaluated.

The risk of reidentification of individuals in the released datasets is another theme in the reviewed articles. A consensus in the literature is that removing personally identifiable information from datasets and aggregating raw data to summary tables does not prevent the reidentification risks. Traditional statistical disclosure limitation techniques and modern privacy models, such as k-anonymity and differential privacy have frequently been employed to de-identify OGD. The literature does not show a survey of de-identification practices in governments' OGD programs. Such a survey could reveal the actual level of reidentification risks of published datasets and the differences among government bodies in their technological and organizational competencies regarding privacy protection.

A promising stream of research is developing frameworks and providing guidelines for performing privacy risk analysis by OGD programs. Governments can use privacy risk frameworks, such as the NIST Privacy Framework (National Institute of Standards and Technology, 2020a) to integrate privacy OGD privacy risks into their organizational risk management practices. However, given the nuances of OGD and the diverse needs of governments at different levels, specialized OGD privacy risk frameworks and guidance are needed.

An excellent example of risk assessment guidance for cities' data-sharing processes is provided by Green et al. (2017). This stream of research has the potential for future research in providing empirical evidence of the probability and consequences of de-identification risks for different types of government data. For example, a significant gap in the literature is the lack of empirical evidence about how OGD is being used as complementary data, to facilitate reidentification, profiling, and advertising through big data analysis by private and public bodies.

Lastly, another gap in the literature, concerns the discriminatory effects of OGD utilization on individuals. There is a need to understand how the social and economic benefits of OGD are distributed among different populations. Moreover, a more relevant question to this review is how privacy disclosure risks are distributed. Knowing the answers to these questions could inform policymakers and OGD practitioners in their efforts to optimize public value.

## About the Author

*Mehdi Barati*

Mehdi Barati is a Ph.D. candidate in information science at the State University of New York at Albany, New York, United States. He conducts research at the intersection of data analytics and work organization, focusing on the use of artificial intelligence in organizations for workforce management.

**Appendix 1**

| Article | Data, Methods, Research Question | Identified Stakeholders | Privacy-related Barriers and Challenges |
|---|---|---|---|
| (Hardy & Maurushat, 2017) | Data: Australian OGD practices<br><br>Methods: a case study<br><br>Research Question:<br><br>Descriptive analysis of Australian government open data practices | OGD programs<br><br>Public agencies<br><br>Citizens | Government agencies usually lack enough expertise for proper de-identification of their data.<br><br>Legal requirements for personal information protection are inherently vague. For example, reasonably identifiable data must not be released. This opacity justifies for agencies to avoid data release. |
| (Henriksen-Bulmer et al., 2019) | Method: conceptual framework using Unified Modelling Language (UML) and evaluation by a case study.<br><br>Research question: Using the contextual integrity framework (Nissenbaum, 2010), this article proposes a meta-model to assess public agencies' privacy risks of releasing datasets. | The public agency that produces the data<br><br>The public agency that shares the data<br><br>Data subjects<br><br>Open data users | Public agencies tend not to publish their dataset with no formal privacy assessment framework to avoid privacy violation consequences.<br><br>Public sector practitioners are unclear on preserving privacy while publishing open data.<br><br>OGD publishers often publish data collected, processed, and stored by other agencies, and they have no control over them. |
| (J. S. Lee & Jun, 2021) | Data: the Korean Innovation Survey (KIS) and the Korea Enterprise Data (KED)<br><br>Methods: a privacy-preserving data mining method for record-linkage of OGD datasets that makes a balance | OGD programs<br><br>Citizens<br><br>Private enterprizes<br><br>Government | |

| | between linkage and privacy disclosure.  Research Question: Developing an algorithm for mining anonymized and already distributed OGD datasets that support heterogeneous data mining for in-depth analysis | | |
|---|---|---|---|
| (Kjaergaard et al., 2020) | This article reviews the current practices and tools of using open data to research occupant-centric design and operation of buildings. | Researchers  Data collectors  Data subjects | |
| (Janssen & van den Hoven, 2015) | Proposes a conceptualizing framework for transparency and privacy in the context of OGD. | Public organizations.  Civil society  citizens | Barriers to opening government data include the creation of information silos, information architectures, and the accompanying privacy frameworks that allow the information release only when there is consent from the data subject or a government decision. |
| (Henninger, 2018) | Data: open government datasets, annual government reports, government agencies' information disclosure logs, and two requests for information and associated correspondence  Methods: case studies  Research Question: What are the relevant trends and the tensions of | Journalists who are seeking personal information through freedom of information act.  Civil society activists  government | The government uses personal privacy as an excuse for refusing freedom of information requests or for releasing government data.  The government might use freedom of information procedures as an obfuscating mechanism for maintaining secrecy while providing "a veil of legitimacy." |

| | conditional exemptions of FOI and OGD for privacy reasons | | |
|---|---|---|---|
| (Beauvais et al., 2021) | This article discusses legal and ethical issues of the open science paradigm and sharing neuroimaging data in open data repositories (including consent and information privacy). | | |
| (Moustaka et al., 2019) | Describes privacy and security concerns of online social network (OSN) data in the context of smart city (SC). | Local authorities<br><br>Citizens<br><br>OSNs | |
| (Piao et al., 2019) | Proposes a framework for publishing open government data in the form of MaxDiff histogram based on differential privacy. | | |
| (Diallo et al., 2021) | Proposes an agent-based simulation method for reconstructing location data of OGD datasets without the identification of individuals. | | |
| (Kao et al., 2017) | Methods: Design and usability study.<br><br>Data: Open datasets from OGD portals<br><br>Research Question: Can a visualization tool help data owners identify and mitigate re-identification | Data owner<br><br>Data subjects | |
| | risks of their data before release? | | |
| (van Zoonen, 2016) | This article provides a framework for understanding and categorizing peoples' privacy concerns in smart cities. Their framework is a two-dimensional space for smart city privacy concerns that consists of the type of data collected and the purpose of data usage. | Citizens<br><br>visitors<br><br>Policymakers<br><br>Local governments | Peoples' privacy concerns and perceptions may not be consistent and predictable. They change temporally and are highly contextual. Therefore, there should be a constant assessment of privacy concerns before opening up any city-data.<br><br>Careful consideration of privacy regulations must be taken. |
| (van Loenen et al., 2016) | | | It is not clear whether and in what context mapping data should be considered personal data and be its release be restricted in open data portals.<br><br>If mapping data is considered personal, then:<br><br>In collection time, the purpose of data processing must be specified, explicit, and legitimate, which is in contrast with open data principles.<br><br>Data subjects could request their data (e.g., their building aerial image) be deleted or modified. |
| (Khan et al., 2017) | Used a mixed-method approach to identify privacy and security issues in smart cities and to propose a security and privacy-aware framework for service | Open data app developers<br><br>Domain experts<br><br>Standard governing bodies | |
| | provisioning in smart cities (SSServProv). | | |
| (Rantala et al., 2020) | Uses qualitative content analysis of public comments on a law proposal to extract major concerns and debates relating to the opening of forest information in Finland. | Forest owners NGOs Government agencies | Ownership of forest data is debated. The government argues that any publicly funded data should be public, but forest owners disagree. The right to erase data from public datasets by forest owners could limit the cohesion and quality of open data. The right to erase public data is considered contrary to open data principles. Privacy arguments are used as a cover for the underlying economic interests of the stakeholders. |
| (Sánchez et al., 2016) | This article combines differential privacy and k-anonymity mechanisms to provide an algorithm for privacy-preserving data publishing, preserving data utility. This method exploits the advantages of k-anonymity (low information loss and lack of assumptions on data uses) and e-differential privacy (robust privacy guarantees). | | |
| (Altman et al., 2016) | This study offers a framework for assessing the risks, threats, and vulnerabilities of personal information privacy in governments' release of | | |

| | personal data. They identify common risks and controls in different stages of the data lifecycle and suggest proper privacy controls, which creates a balance between privacy and utility of the government data. | | |
|---|---|---|---|
| (Borgesius et al., 2015) | This article is a law review article.<br><br>-Research question: How privacy and related interests can be respected without hampering the benefits of releasing government data as OGD? | Public<br><br>Individuals<br><br>Public sector<br><br>Private institutions<br><br>Marketing agencies<br><br>Researchers | |
| (Yazdanie & Orehounig, 2021) | -Literature review<br><br>- What are the technical, methodological & institutional gaps in energy planning models? | | |
| (Fredj et al., 2015) | Existing generalization algorithms and experiment /<br><br>Abstraction process: providing simplified representation of algorithms. /<br><br>How does a data publisher choose an anonymization technique and a proper algorithm for its implementation? | Open data publishers<br><br>Algorithm developers | |

　　　　　　　　

| | | | |
|---|---|---|---|
| (Hasanzadeh et al., 2020) | Data: The data was collected using an online PPGIS method that combines Internet maps with traditional questionnaires (N=884)/ Method: A customized bimodal Gaussian displacement algorithm coupled with donut anonymization/ What are the privacy concerns of PPGIS data? Develop a practical PPGIS data anonymization approach and strategy. How can PPGIS data anonymization affect data quality? | | |
| (Scassa & Conroy, 2017) | Method: Report review Research Question: What strategies can be used to balance information privacy and transparency goals in the release of government information? | | -License restrictions Moreover, technological barriers are in contrast with open data principles. -These barriers do not protect against privacy violations in practice. |
| (Gkoulalas-Divanis & Mac Aonghusa, 2014) | Method: Case study and review report, QuerioCity is an open urban information management platform that is based on semantic technologies to capture, manage, interconnect and enable the consumption of urban | | |

| | open data to provide insight into the operations of a city. Research Question: What is common privacy-preserving data publishing in the context of open information management platforms? | | |
|---|---|---|---|
| (Sinnott et al., 2016) | Data: a survey of 25,000+ Victorians by the Department of Health in Victoria Method: A case study Research Question: providing geospatial unit-level data to researchers in an open government data platform without revealing personal information and the location of individual respondents. | Department of health Urban researchers | |
| (Rohunen et al., 2014) | Data: stakeholder interviews, user interviews, and a user survey from two pilot studies. (N=10 for interview, and N=62 for survey) Methods: Semi-structured theme Interviews and surveys. Research Question: What are the main factors influencing the | Concerns about what purposes the data would be used. concern about those to whom their data would be disclosed. concerns about what purposes the data are used for | |
| | individuals' willingness to disclose their driving data? | | |
| (Conroy & Scassa, 2015) | Method: critical review and case study<br><br>Research Question: How can a balance be struck between personal privacy and government transparency in the context of OGD and proactive disclosure? | | The risk of re-identification should/not be evaluated in the context of the data to be released. Future developments in technology and other available data should/not be considered in evaluating the risks.<br><br>In an open data context, the balance is also between privacy and the economic value of the data, and transparency is less important. |
| (Meijer et al., 2014) | Data: interview and survey of stakeholders<br><br>Method: literature review and case study<br><br>Research Question: What are the contradicting values of open government data, and can they be reconciled? | | |
| (Rubinstein, 2018) | Method: Law review and case studies<br><br>Research Question:<br><br>How privacy and open data regulation at the level of local governments can balance open data, information privacy, and related public values. | | |
| (Green et al., 2017) | | Cities<br><br>Developers | |
| | | Citizens<br><br>Researchers<br><br>Data subjects<br><br>Communities | |
| (Walsh et al., 2018) | Method: Literature review and case study<br><br>Question: What are the common privacy risks and de-identification methods in sharing psychological and psychiatry clinical data? | Patients<br><br>Researchers<br><br>Policymakers<br><br>IRBs | |
| (Michener & Ritter, 2017) | Method: Semi-structured interviews<br><br>Data: Interviews with 60 stakeholders in Brazil and the UK | Students<br><br>Parents<br><br>School administrators<br><br>politicians | |
| (Wang et al., 2019) | Method: exploratory analysis of usage data and thematic analysis interviews<br><br>Data: usage statistics of OGD portal and interviews with junior managers | Open data activists<br><br>Information activists<br><br>Administrators | Legal compliance with GDPR and national and local privacy laws.<br><br>The concerns of diminishing public trust and reputation of agencies in case of privacy violation discourage managers from data release. |
| (Heijlen & Crompvoets, 2021) | Method: Literature review and ecosystem mapping.<br><br>Data: | Health data providers<br><br>Health data producers<br><br>Health data users | Legal barriers of data protection laws.<br><br>Data quality barriers (de-identification and aggregation reduce data quality for research and other uses) |

|  |  | Service developers<br><br>Pharmaceuticals<br><br>Insurance providers |  |
| --- | --- | --- | --- |
| (Badu-Marfo et al., 2019) | Method: Experimental Simulations.<br><br>Data: Home locations of 7,985 respondents from a large-scale smartphone travel survey conducted by the City of Montreal using the app MTL Trajet developed by the Concordia University TRIP Lab<br><br>Research Question: How do the two location privacy protection algorithms (Donut Geo-mask and Geo-Indistinguishability) perform in privacy and utility protection. |  |  |
| (Jeon et al., 2021) | Method: Experimental Simulations.<br><br>Data: OGD RDF data<br><br>Research Question: Integrate the l-diversity anatomy de-identification method with existing k-anonymity methods to improve the privacy protection of RDF data published in OGD portals. |  |  |
| (Shaham et al., 2021) | Method: Experimental Simulations.<br><br>Data: location trajectories and spatiotemporal trajectories published as OGD.<br><br>Research Question: proposing a robust framework for the anonymization of spatiotemporal trajectory datasets termed machine learning-based anonymization (MLA) | | |
| (B. Lee et al., 2021) | Method: used privacy heuristics, available guidance, and codes of practice to develop Data Sharing Privacy Review Procedures.<br><br>Data: Covid-19 patient-level data from all jurisdictions of US sent to CDC<br><br>Research Question: creating privacy-protected public datasets from CDC's Covid-19 patient-level data | Patients<br><br>Researchers<br><br>Data providers (jurisdictions)<br><br>Data publisher (CDC) | |
| (Luthfi et al., 2018) | Method: an explanatory model to create a Bayesian-belief Network of the risks of opening government data<br><br>Data: health patient records | | |

|  | Research Question: |  |  |
|---|---|---|---|
| (Vasileva et al., 2018) | Method: thematic analysis and quantitative analysis.<br><br>Data: data were collected from 23 interviews with key stakeholders and a survey with 205 responses.<br><br>Research Question: Could smart campus projects set examples for smart city initiatives regarding open data utilization and challenges? | Students<br><br>Faculty and staff<br><br>Administrators<br><br>Private sector service providers | Privacy and security of the data are primary challenges mentioned by respondents on campus. However, stakeholders' perception of risk could be mitigated with data protection and anonymization measures, appropriate communication of the protection measures, and data sharing benefits.<br><br>Upgrading and interconnecting information systems needed for data collection and management could also reduce the concerns. |
| (El Emam et al., 2012) | Method: a risk-based approach to identifying re-identification attack scenarios and propose proper de-identification algorithms.<br><br>Data: longitudinal public health dataset in the Heritage Health Prize (HHP) context.<br><br>Research Question: how to de-identify the HHP dataset concerning plausible attack scenarios and corresponding risks. | Researchers<br><br>Government agency<br><br>Health data subjects (patients) |  |
| (Austin & Lie, 2019) | Method: Conceptual framework<br><br>Data: Three case studies<br><br>Research Question:<br><br>Is there an alternative framework for sharing public sector data with interested parties and preserving privacy other than de-identification and releasing non-PII? | Public sector agencies<br><br>Private sector vendors<br><br>citizens | |
| (Scassa, 2014) | Method: conceptual<br><br>Data:<br><br>Research Question:<br><br>What are the challenges of striking a balance between information privacy and transparency in OGD programs? | Government<br><br>Citizens<br><br>OGD programs | The difficulty of control of information dissemination in the OGD context.<br><br>Finding a proper balance between privacy and transparency and accountability is difficult.<br><br>The blurring line between the data collected by the private sector and the data collected by government bodies in cases where the private sector acts as an intermediary or service provider. |
| (Altman et al., 2018) | Method: review and discussion<br><br>Data:<br><br>Research Question: what lessons can be learned from privacy risks and protection approaches of longitudinal research studies to understand and mitigate privacy risks of OGD programs? | Health providers<br>Patients<br>Government | The high dimensionality of big data creates challenges that traditional technical and procedural privacy controls cannot address. Personal information obtained from unstructured data, including text, audio, and video, cannot be limited to de-identification. |
| (Piao et al., 2017) | Method: conceptual framework and data experiment<br><br>Data:<br><br>Research Question: developing a framework for publishing government data based on a differential privacy model in two interactive and non-interactive modes. | Government<br><br>citizens | |
| (Huang et al., 2019) | Method: knowledge discovery in the database procedure<br><br>Data: de-identified Electronic Toll Collection from Taiwan<br><br>Research Question: is re-identification of de-identified ETC open data possible? Which de-identification methods are more robust? | Passenger<br><br>developers | |
| (Lavrenovs & Podins, 2016) | Method: using public information, including public transportation stops combined with OGD, to re-identify individual passengers.<br><br>Data: Public transportation e-talons ride registration data | Public transport passenger<br><br>Attackers<br><br>Public agencies<br><br>developers | |

| | from Riga municipality open government portal<br><br>Research Question: what privacy attacks can run on public transportation data published on OGD? | | |
|---|---|---|---|